\documentclass[aps,12pt,superscriptaddress,tightenlines]{revtex4}
\usepackage{amsmath,amssymb,amsthm,graphicx,bm,color}

\renewcommand{\Delta}{\varDelta} 
\renewcommand{\Gamma}{\varGamma} 
\renewcommand{\Omega}{\varOmega} 
\renewcommand{\Phi}{\varPhi} 
\renewcommand{\Psi}{\varPsi} 
\renewcommand{\Sigma}{\varSigma} 
\renewcommand{\Theta}{\varTheta} 
\renewcommand{\epsilon}{\varepsilon}

\newcommand{\Z}{\mathbb{Z}}

\begin{document}

\title{A proposal for detecting second order topological quantum
phase}

\author{Roman V. Buniy}

\email{roman@uoregon.edu}

\affiliation{Institute of Theoretical Science, University of Oregon,
Eugene, OR 94703}

\author{Thomas W. Kephart}

\email{tom.kephart@gmail.com}

\affiliation{Department of Physics and Astronomy, Vanderbilt
University, Nashville, TN 37235} 

\date{\today}

\begin{abstract}
 Gaussian linking of a semiclassical path of a charged particle with a
  magnetic flux tube is responsible for the Aharonov-Bohm effect,
  where one observes interference proportional to the magnitude of the
  enclosed flux. We construct quantum mechanical wave functions where
  semiclassical paths can have second order linking to two magnetic
  flux tubes, and show there is interference proportional to the
  product of the two fluxes.
\end{abstract}

\pacs{}

\maketitle

Topological phases can arise when a particle traverses semiclassical
paths that cannot be deformed into each other due to some obstruction
in an experimental setup, for example, paths that pass on opposite
sides of an infinitely long solenoid. If the particle is charged, and
there is a magnetic flux confined within the obstruction, then the two
paths experience different vector potentials. This generates a phase
difference for the two topologically different paths and causes
interference when the particle is detected. The magnitude of the phase
is a measure of the Gaussian linking of the particle path with the
solenoid. What we have described here is the Aharonov-Bohm effect
\cite{Aharonov:1959fk}. But, this is not the full story, as we will
now argue.

\begin{figure}[ht]
\includegraphics[width=9cm]{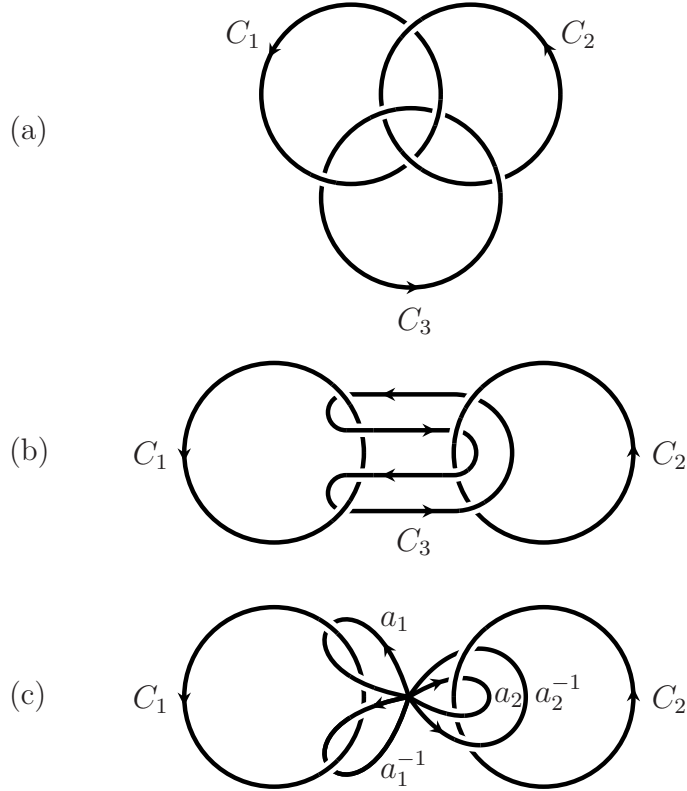}
\caption{\label{figure-commutator} (a) The Borromean rings. The
  topological significance of this arrangement is that while no pair
  of loops is linked, the triple of loops is linked. This is the
  simplest configuration with zero first order (Gaussian) linking and
  nonzero second order linking. (b) To obtain this figure from (a),
  pull $C_1$ and $C_2$ apart for flexible $C_3$. (c) To obtain this
  figure from (b), pinch $C_3$ at the point $x_0$ so that it becomes a
  commutator loop $C_3=a_1a_2a_1^{-1}a_2^{-1}$.}
\end{figure}

\begin{figure}[ht]
\includegraphics[width=6.5cm]{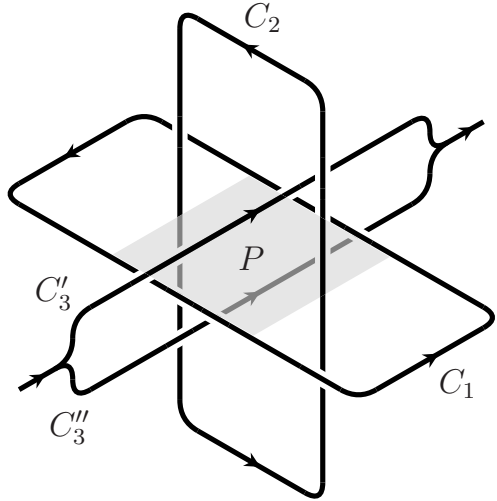}
\caption{\label{figure-experiment} Shown is a schematic of a Borromean
ring arrangement to detect the second order phase $\phi_{12}$, where
$C_1$ and $C_2$ are magnetic solenoids (leads not shown) carring flux
$\Phi_1$ and $\Phi_2$, and $C_3'$ and $C_3''$ correspond to two
topologically distinct paths and are parts of the closed path
$C_3=C'_3{C''_3}^{-1}$ of a charged particle path starting from the
source and ending at the screen. To prevent second order (gaussian)
linking of the wave function with the solenoids one would install a
rectangular plate $P$ in the plane of $C_1$ that covers the area
between the two sides of $C_1$ and fills the region between the sides
of $C_2$. For particle wave packets that do not spread much beyond the
center of the region containing the plate, only second order linking
will be detected at the screen. }
\end{figure}

Higher order linking is possible. Consider the Borromean rings, an
arrangement of three loops inextricably linked~\cite{Rolfsen} but with
no first order (Gaussian) linking between any pair (see
Fig.~\ref{figure-commutator}a).
To see the higher order linking in more detail, we let ring $C_3$ be
flexible and pull rings $C_1$ and $C_2$ apart while keeping their
shapes fixed. This gives Fig.~\ref{figure-commutator}b. Next, we pinch
the lines of ring $C_3$ in Fig.~\ref{figure-commutator}b together at
point $x_0$ to form Fig.~\ref{figure-commutator}c~\cite{Hatcher}. Now
we follow the semiclassical path of ring $C_3$ to see how it is linked
with rings $C_1$ and $C_2$. From Fig.~\ref{figure-commutator}c we see
that we get four components: $a_1$, followed by $a_2$, followed by
$a_1^{-1}$, and then by $a_2^{-1}$. Here $a_1$ links through $C_1$ and
$a_2$ links through $C_2$ in the positive sense respectively, while
$a_1^{-1}$ and $a_2^{-1}$ link through $C_1$ and $C_2$ in the negative
sense \cite{pi1}. So the entire path $C_3$ runs through $C_1$ once in
the positive and once in the negative sense for a total Gaussian
linking of zero with $C_1$. Likewise, there is no Gaussian linking
with $C_2$. But the total path $C_3$ is not trivial. This is because
the paths $a_1$ and $a_2$ do not commute. In fact, $C_3$ is just the
commutator, which we can write in multiplicative form as
$C_3=a_1a_2a_1^{-1}a_2^{-1}$. It is this commutator that leads to a
new phase. To see this, we must introduce a physical system that
displays the properties we have been describing. We need the topology
of Fig.~\ref{figure-commutator}, but in such an arrangement that loop
$C_3$ corresponds to the path of a particle and loops $C_1$ and $C_2$
to solenoids. This should not be difficult to arrange experimentally,
and a sketch is provided in Fig.~\ref{figure-experiment}.

In the Aharonov-Bohm case, the wave function along a path $\Gamma$ can
be written as $\psi({\bm A})=\psi(0)\exp{({i\int_{\Gamma} {\bm A\cdot}
d{\bm x}})}$, where we are using natural units $\hbar=c=1$ with unit
charge to simplify the analysis, but will restore physical units when
we reach our results. The interference between wave amplitudes from
the two semiclassical paths $C'$ and $C''$ around a closed path
$C=C'{C''}^{-1}$ is
\begin{equation}
  \psi({\bm A}) =\psi'({\bm A})+\psi''({\bm
    A})=e^{i\beta}\left[\psi'(0) +e^{i\phi}\psi''(0)\right],
\end{equation}
where $\beta$ is an overall irrelevant phase and the important
relative phase is
\begin{equation}
  \phi=\oint_{C=\partial S} {\bm A} \cdot d{\bm x }=\int_S {\bm B}
    \cdot d{\bm S}=\frac{e\Phi}{\hbar c}.
\end{equation}
 in SI units, where $\Phi$ is the magnetic
flux enclosed in the solenoid.

Now let us return to the Borromean ring configuration and follow the
semiclassical path of a particle around the circuit $C_3$ where we now
take $C_1$ and $C_2$ to be a pair of unlinked solenoids. If, and only
if, the particle path has no net first order linking with either
solenoid $C_1$ or $C_2$, will we then define a gauge \cite{Massey},
\cite{Berger} that describes the higher order linking \cite{gauge
comment}. That gauge is
\begin{equation}
  {\bm A_{12}} =\tfrac{1}{2}(\gamma_1{\bm A_2}-\gamma_2{\bm A_1}),
  \label{A12}
\end{equation}
where subscripts 1 and 2 refer to the solenoids along $C_1$ and $C_2$,
and
\begin{equation}
\gamma_k=\delta_k + \int_{\Gamma} {\bm A_k\cdot }d{\bm x}. 
\end{equation}
Here ${\bm A_1}$ and ${\bm A_2}$ are the vector potentials due to the
two solenoids, and $\Gamma$ is the path that will run along $C_3$.

We now want to calculate the overall phase difference $K^{-1} \phi
_{12}= \int_{C_3} {\bm A}_{12}\cdot d{\bm x}$. (The $K^{-1}$
normalization factor multiplying $\phi _{12}$ will be discussed
below.) Table~\ref{table} follows the path step by step through the
experimental setup along path $C_3$ using the gauge ${\bm A_{12}}$.
\begin{table*}
  \caption{\label{table}Phase components along the path $C_3$.}
  \begin{ruledtabular}
    \begin{tabular}{cccc}
      path segment $\Gamma$ & $\gamma _{1}$ & $\gamma _{2}$ & $K^{-1}\phi
      _{12}(\Gamma)$ \\ \hline $1$ & $\delta _{1}$ & $\delta _{2}$ &
      $0$ \\ $a_1$ & $\delta _{1}+\Phi _{1}$ & $\delta _{2}$ &
      $-\frac{1}{2}\Phi _{1}\delta _{2}$ \\ $a_1a_2$ & $\delta _{1}+\Phi
      _{1}$ & $\delta _{2}+\Phi _{2}$ & $\frac{1}{2} [(\delta
      _{1}+\Phi _{1})\Phi _{2}-\Phi _{1}\delta _{2}]$ \\ $a_1a_2a_1^{-1}$ &
      $\delta _{1}$ & $\delta _{2}+\Phi _{2}$ & $\frac{1}{2}[(\delta
      _{1}+\Phi _{1})\Phi _{2}-\Phi _{1}\delta _{2}+\Phi _{1}(\delta
      _{2}+\Phi _{2})]$ \\ $C_{3}=a_1a_2a_1^{-1}a_2^{-1}$ & $\delta _{1}$ &
      $\delta _{2}$ & $\Phi _{1}\Phi _{2}$
    \end{tabular}
  \end{ruledtabular}
\end{table*}
The first column labels the current positions on the path, the next
two columns are the cumulative values of $\gamma_1$ and $\gamma_2$ at
these points, and the last column gives the value of $\phi
_{12}(\Gamma)$ at these points.  In the third row we have used $a_1$
to take us from $x_0$ around $C_1$ and back to $x_0$. In the process
$\gamma_1$ has increased by $\Phi_1$ since $\int_{\Gamma=a_1} {\bm
A_1}\cdot d{\bm x} =\Phi_1$ while $\gamma_2$ stays fixed since $
\int_{\Gamma=a_1} {\bm A_2\cdot }d{\bm x}=0$. Hence we have
\begin{align}
K^{-1} \phi_{12}(a_1) =\tfrac{1}{2}\gamma_1 \int_{\Gamma=a_1} {\bm
A_2\cdot }d{\bm x}- \tfrac{1}{2}\gamma_2\int_{\Gamma=a_1} {\bm
A_1\cdot }d{\bm x} =-\tfrac{1}{2}\Phi_1\delta_2.
\end{align}
From here it is obvious how to generate
the remaining entries in the table.
\begin{figure}[ht]
\includegraphics[width=11cm]{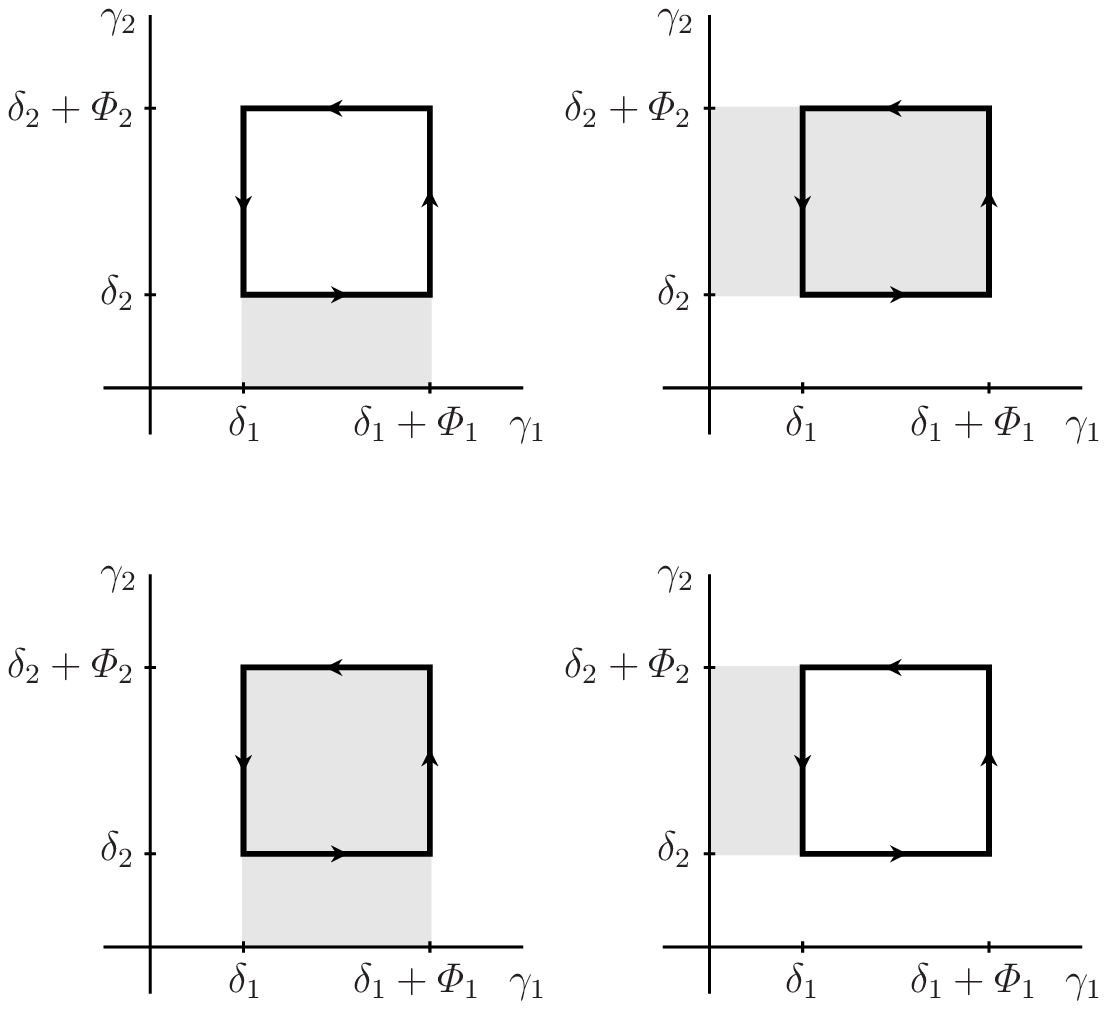}
\caption{\label{figure-commutator-steps} Graphic representation of the
information in Table I.}
\end{figure}
An alternative representation of this information is given in
Fig.~\ref{figure-commutator-steps}. Here the path $C_3$ begins at the
initial position $(\delta_1,\delta_2)$. We first use $a_1$ to travel
to $(\delta_1+\Phi_1,\delta_2)$, picking up an area $\delta_2\Phi_1$,
which corresponds to a contribution of $-\frac{1}{2}\delta_2\Phi_1$ to
$K^{-1} \phi_{12}$ (see Eq.~(\ref{A12})). Next, $a_2$ takes us to
$(\delta_1+\Phi_1,\delta_2+\Phi_2)$ and it generates a contribution
$\frac{1}{2}(\delta_1+\Phi_1)\Phi_2$. Next, $a_1^{-1}$ takes us to
$(\delta_1,\delta_2+\Phi_2)$ and contributes
$\frac{1}{2}\Phi_1(\delta_2+\Phi_2)$. Finally, $a_2^{-1}$ returns us
to $(\delta_1,\delta_2)$ and contributes $-\frac{1}{2}\delta_1\Phi_2$,
for a total phase of $K^{-1} \phi_{12}=\Phi_1\Phi_2$ for traversing
the full loop $C_3$. The last row of Table~\ref{table} (or the full
loop in Fig.~\ref{figure-commutator}c) gives the final result for the
full path when $\Gamma=C_3$. We find
\begin{equation}
  \phi _{12}(C_3)= K\frac{e^2}{\hbar^2c^2}\Phi _{1}\Phi _{2}
  \label{phi12}
\end{equation}
once physical units have been restored. Figure~\ref{figure-experiment}
provides a schematic of the Borromean ring experimental setup with two
solenoidal rings and split charged particle path.

Equation~(\ref{phi12}) is our main result and may be surprising in
several respects. First and foremost, $\phi _{12}(C_3) $ does not
vanish, even though the wave function has no first order linking with
either solenoid. Second, the overall phase is proportional to the
product of the fluxes from the two solenoids. Third, the result is not
difficult to generalize to more complicated paths with multiple second
order linking as we will show below, and to higher order of linking as
we will show elsewhere \cite{BK}.

Before proceeding let us finally discuss normalization factor in the
phase. Recall Dirac's magnetic monopole requires a string (return flux
tube).  The string can be made unobservable if it carries an integer
number of flux quanta.  Likewise the Aharonov-Bohm phase is
unobservable if the phase shift is a multiple of $2\pi$, and the
magnetic flux enclosed by the particle paths is an integer multiple of
the flux quantum. In Fig.~\ref{figure-experiment} we make a similar
requirement.  If both $C_1$ and $C_2$ carry quantized flux, i.e., if
both $\frac{e\Phi_1}{\hbar c}$ and $\frac{e\Phi_2}{\hbar c}$ are
integer multiples of $2\pi$, then we expect the second order linking
to be unobservable \cite{2pi}. This is the case if we include the $K=
\frac{1}{2\pi}$ normalization factor \cite{2pi2} in Eq.~(\ref{phi12}).

\begin{figure}[ht]
\includegraphics[width=8cm]{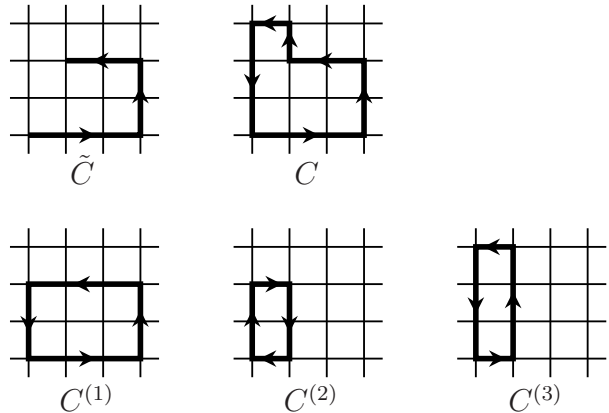}
\caption{\label{figure-lattice} Paths in lattice space generated by
$a_1$ and $a_2$. Open paths have first order linking, closed paths do
not.}
\end{figure}

Before concluding let us explore the case of multiple second order
linking.  Again, consider two unlinked closed rings $C_1$ and $C_2$
and a third path $C$ that will wrap around them. $C$ starts at point
$x_0$ and then wraps via subpaths $a_1$ and $a_2$ some number of
times. We define a lattice space of paths where $a_1$, $a_2$,
$a_1^{-1}$ and $a_2^{-1}$ are right, up, left, and down steps by one
lattice spacing, respectively. For example, consider the first frame
in Fig.~\ref{figure-lattice}, where $\tilde C=a_1^3a_2^2a_1^{-2}$. The
accumulated first order linking $\tilde\phi_k$ is the total number of
times $C$ wraps around $C_k$ ($k=1,2$), i.e., the projected distance
on the $k$-axis from the starting point. Here $\tilde\phi_1=1$ and
$\tilde\phi_2=2$. But notice the path is not closed, and so ${\bm
A_{12}}$ cannot be defined, and there can be no second order
linking. Next note that any closed path has no net first order
linking, i.e., the net numbers of horizontal and vertical moves are
both zero, but this is just when we can define ${\bm A_{12}}$. Now
consider the closed path in the second frame of
Fig.~\ref{figure-lattice}, where
$C=a_1^3a_2^2a_1^{-2}a_2a_1^{-1}a_2^{-3}$. This path can be written as
a product of commutators, $C=C^{(1)}C^{(2)}C^{(3)}$, where the
commutators are themselves closed paths (see the bottom row of
Fig.~\ref{figure-lattice}). The total accumulated phase for $C$ is the
sum of those for $C^{(1)}$, $C^{(2)}$, $C^{(3)}$; in this case,
$\phi_{12}= 6-2+3=7$, and this corresponds to the total number of
cells of the lattice enclosed by path $C$. (Recall that the simple
Borromean ring commutator is $a_1a_2a_1^{-1}a_2^{-1}$, which encloses
one lattice cell.) The result for enclosed flux by an arbitrary closed
path $C$ is
\begin{align}
  \phi_{12}=nK\frac{e^2}{\hbar^2 c^2}\Phi_1\Phi_2,
\end{align} 
where $n$ is the number of cells enclosed by the path.

In summary, first order (Gaussian) linking leads to interference with
phase proportional to enclosed flux in the case of the Aharonov-Bohm
effect. Higher order linking also leads to interference, but with
phases proportional to products of fluxes from different
solenoids. Even though path components $a_1$ and $a_2$ in the above
example do not commute, the phase is still abelian as required
\cite{path-integral}. Our analysis needs nothing more than quantum
mechanics and a judicious choice of gauge, and our conclusions are
easily testable with tabletop experiments using known techniques.

\begin{acknowledgments}
  We thank Jason Cantarella for a useful discussion and for pointing
  out Ref.~\cite{Berger}. TWK thanks the Aspen Center for Physics for
  hospitality while this work was in progress. The work of RVB was
  supported by DOE grant number DE-FG06-85ER40224 and that of TWK by
  DOE grant number DE-FG05-85ER40226.
\end{acknowledgments}


\begin{thebibliography}{100}
\bibitem{pi1}$a_1$ and $a_2$ are generators of the fundamental group
  $\pi_1(M)$ of the space $M=R^3-(C_1\cup C_2)$. If $C_1$ and $C_2$
  are unlinked, then $\pi_1(M)=\Z*\Z$, where $*$ indicates the free
  product \cite{free-groups}. Paths with vanishing first order linking
  correspond to elements of the commutator subgroup of $\pi_1(M)$.
\bibitem{gauge comment} $a_1$ is a closed path for the (first order)
  choice of gauge used in the Aharonov-Bohm case, but it is only part
  of the path $C_3$.  At second order, a closed path is a commutator
  of the generators of closed paths at first order.  When the phase
  depends on the choice of the point $x_0$ and we do not have a closed
  path, we are not measuring interference since we have not recombined
  the two halves of the wave function. So one can argue that if we are
  looking for interference, then we should look for a gauge where we
  have closed paths, and $a_1$, $a_1a_2$, and $a_1a_2a_1^{-1}$ are not
  closed for $A_{12}$. Since for the AB gauge choice $a_1$, $a_1a_2$,
  and $a_1a_2a_1^{-1}$ deliver interference, while in $A_{12}$ they
  give results dependent on the location of the point $x_0$, we see
  $A_{12}$ may be an allowed gauge choice, but it is not a good choice
  for these paths. It only becomes a good choice when we are looking
  at the full commutator path.
\bibitem{2pi2} Similarly, we expect higher order cases where the
  phases are proportional to the product of $n$ fluxes to have
  normalization factors $(2\pi)^{1-n}$ as will be discussed in
  \cite{BK}.
\end{thebibliography}
\end{document}